\documentstyle[12pt]{article}
\def\thefootnote{\fnsymbol{footnote}}

\newcommand{\sts}{\footnotesize}
\newcommand{\scr}{\scriptsize}

\newcommand{\spz}{\hspace{0.7cm}}
\newcommand{\sps}{\hspace{3mm}}

\newcommand{\fr}{\rightarrow}
\newcommand{\de}{\partial}
\newcommand{\ri}{\right}
\newcommand{\lf}{\left}
\newcommand{\th}{\theta}

\newcommand{\ep}{\varepsilon}
\newcommand{\eq}{\begin{equation}}
\newcommand{\en}{\end{equation}}
\newcommand{\bea}{\begin{eqnarray}}
\newcommand{\eea}{\end{eqnarray}}
\newcommand{\acc}{\\[3mm]}
\newcommand{\ba}{\begin{array}}
\newcommand{\ea}{\end{array}}
\newcommand{\ds}{\displaystyle}
\newcommand{\ZZ}{\hbox{{\rm Z{\hbox to 3pt{\hss\rm Z}}}}}
\newcommand{\Y}{\Upsilon}

\newcommand{\CZ}{{\cal{Z}}}

\newcommand{\virg}{\spz,}
\newcommand{\pu}{\spz.}

\newcommand{\CS}{{\cal S}}

\newcommand{\D}{{\cal D}}

\newcommand{\N}{{\cal N}}
\newcommand{\NP}[1]{Nucl.\ Phys.\ {\bf #1}}
\newcommand{\PL}[1]{Phys.\ Lett.\ {\bf #1}}

\newcommand{\PR}[1]{Phys.\ Rev.\ {\bf #1}}

\newcommand{\IJMP}[1]{Int.\ J.\ Mod.\ Phys.\ {\bf #1}}

\newcommand{\PTP}[1]{Prog.of \ Theo.\ Phys.\ {\bf #1}}

\newcommand{\JMP}[1]{J.\ Math.\ Phys. \ {\bf #1}}
\newcommand{\PA}[1]{Physica  {\bf #1}}
\newsavebox{\Bn}
\sbox{\Bn}{\begin{picture}(52,5)(0,-3.5)
\multiput(10,0)(10,0){5}{\circle*{3}}
\put(10,0){\line(1,0){8}}
\put(50,21.5){\line(0,1){7}}
\put(49.5,31.5){\line(0,1){7}}
\put(50.5,31.5){\line(0,1){7}}
\put(50,1.5){\line(0,1){7}}
\multiput(50,0)(0,10){5}{\circle{3}}
\multiput(50,11.5)(0,1){7}{\circle*{.2}}
\multiput(20,0)(1,0){9}{\circle*{.2}}
\put(30,0){\line(1,0){10}}
\put(40,0.5){\line(1,0){10}}
\put(40,-0.5){\line(1,0){10}}
\put(12,-2){\makebox(0,0)[t]{{\protect\scr 1}}}
\put(22,-2){\makebox(0,0)[t]{{\protect\scr 2}}}
\put(32,-2){\makebox(0,0)[t]{{\protect\scr {\em n$_1$}--1}}}
\put(42,-2){\makebox(0,0)[t]{{\protect\scr {\em n$_1$}}}}
\put(52,-2){\makebox(0,0)[t]{{\protect\scr {\em n$_1$ + } 1}}}
\multiput(60,40)(10,0){5}{\circle{3}}
\put(40,10){\makebox(0,0)[t]{{\protect\scr {\em n$_1$ + }  2}}}
\put(35,20){\makebox(0,0)[t]{{\protect\scr {\em n$_1$ + n$_2$}--1}}}
\put(35,30){\makebox(0,0)[t]{{\protect\scr {\em n$_1$ + n$_2$}}}}
\put(35,40){\makebox(0,0)[t]{{\protect\scr {\em n$_1$ + n$_2$}+1}}}
\multiput(51.5,40)(10,0){1}{\line(1,0){7}}
\multiput(61.5,40)(1,0){7}{\circle*{.2}}
\multiput(71.5,40)(10,0){2}{\line(1,0){7}}
\put(91.5,40.5){\line(1,0){7}}
\put(91.5,39.5){\line(1,0){7}}
\multiput(100,38.5)(0,-1){9}{\circle*{.2}}
\put(100,28.5){\circle{3}}
\multiput(101.5,28.5)(1,0){17}{\circle*{.2}}
\put(120,28.5){\circle{3}}
\multiput(120,27)(0,-1){11}{\circle*{.2}}
\put(120,15){\circle{3}}
\put(120.5,6.5){\line(0,1){7}}
\put(119.5,6.5){\line(0,1){7}}
\multiput(120,5)(10,0){4}{\circle{3}}
\put(121.5,5){\line(1,0){7}}
\multiput(131.5,5)(1,0){7}{\circle*{.2}}
\put(141.5,5){\line(1,0){7}}
\put(151.3,5.2){\line(1,1){7}}
\put(151.3,4.8){\line(1,-1){7}}
\put(159.3,-3.8){\circle{3}}
\put(159.3,13.8){\circle{3}}
\put(160,-8.8){\makebox(0,0)[t]{{\protect\scr {\em n$_T \equiv f$}}}}
\put(110,15){\makebox(0,0)[t]{{\protect\scr {\em n$_T$-n$_F$}}}}
\put(160,20.8){\makebox(0,0)[t]{{\protect\scr {\em n$_T$}--1 $\equiv
\bar{f}$}}} \end{picture}}
\begin{document}
\begin{titlepage}
\vskip 0.5cm
\begin{flushright}
DTP/95/16 \\
January, 1995
\end{flushright}
\vskip0.5cm
\begin{center}
{\large {\bf
New functional dilogarithm identities}\\
{\bf  and} \\
{\bf  sine-Gordon Y-systems}}\\
\end{center}
\vskip 0.6cm
\centerline{ R.Tateo}
\vskip 0.6cm
\centerline{\sl  Dipartimento di Fisica
Teorica dell'Universit\`a di Torino\footnote{e-mail: tateo@to.infn.it}}
\centerline{\sl via P.Giuria 1, I--10125 Torino, Italy}
\vskip .2 cm
\centerline{ and }
\centerline{\sl  Department of Mathematics, University of
Durham\footnote{e-mail: roberto.tateo@durham.ac.uk}}
\centerline{\sl South Road,  DH1 3LE  Durham, England}
\vskip 2.cm
\begin{abstract}
\vskip0.2cm
\noindent
The sine-Gordon Y-systems and those of the minimal $M_{p,q}+\phi_{13}$
models  are determined  in a compact form
and a correspondence between  the rational numbers and a new infinite
family of
multi-parameter functional equations for the Rogers dilogarithm is pointed
out. The  relation between the TBA-duality and the massless RG
fluxes in the minimal  models recently conjectured is   briefly discussed.
\end{abstract}
\end{titlepage}
\setcounter{footnote}{0}
\def\thefootnote{\arabic{footnote}}
\newpage
\section{Introduction}
In~\cite{ys} Al.B. Zamolodchikov has
proposed a transformation of the TBA
equations~\cite{yy,al1,km2}
for the so-called $ADET$ diagonal scattering showing
clearly  their
relation with  a system of functional equations now known  as $ADET$
Y-systems. Many other  Y-systems  have now been  discovered, and
the importance of these relations  in connection to the theory of
integrable models  has greatly increased: not only excited states ,
different phases of the same model and models
that differ  only in the  boundary conditions imposed
satisfy the same Y-system equations in all known cases, but  also
many  definitely different models lead to the same
equations~\cite{al2,ma,km3,fee,blz}.
The  origin of this kind of  universality is partially
due to the rigid structure of  this  object, imposed by two remarkable
properties:  the stationary
dilogarithm sum-rules, related to the UV and IR Casimir effect, and the
periodicity related to the
conformal dimension of the UV (IR) perturbing operator.
In a recent paper~\cite{gt} a new important mathematical property
has been pointed out: using as a basic object the so-called
$ADET \diamond ADET$
Y-systems~\cite{rtv,rt} a new family of  multi-parameter functional
dilogarithm~\footnote{In connection with the K-theory, R. de Jeu~\cite{rob}
has also checked that  some of these   Rogers's FDE
have a corresponding m=2 Bloch-Wigner~\cite{don} functional equation.}
 equations (FDE)~\cite{lewin}
generalization of the Abel's and Euler's formulas has been proposed
\footnote{ See~\cite{kir0} for a list of the different  branches of
mathematics and physics in witch the dilogarithm  emerging.}.
\noindent
The above two   properties and the physical interpretation
of this result  given in~\cite{gt} suggest the  possibility of finding new
FDE using the Y-systems of the known models:
the aim of this letter is to  show how to construct in a simple way
FDE  using the sine-Gordon TBA  or its
reduced form
associated to the minimal models.
 The first part of this letter is devoted
to  the explicit formulation of these two families of Y-systems in a compact
form; this is also an original result and the
``fusion at the massive node''~\cite{fein} can be generalized to
these Y-systems for  obtaining   the TBA equations of
all the  $Z_n$ supersymmetric versions~\cite{ber} of the sine-Gordon or
their reduced
models. In the second part we present the result concerning the new
functional dilogarithm identities and  finally  the relation between the
TBA-duality
connecting the theories $M_{p,q} \leftrightarrow M_{q-p,q}, Z_{p-1}
\leftrightarrow
M_{p,p+1}$ and the presence of
non-unitary (unitary)  massless fluxes in the minimal
models is briefly discussed.

\section{The sine-Gordon Y-systems}
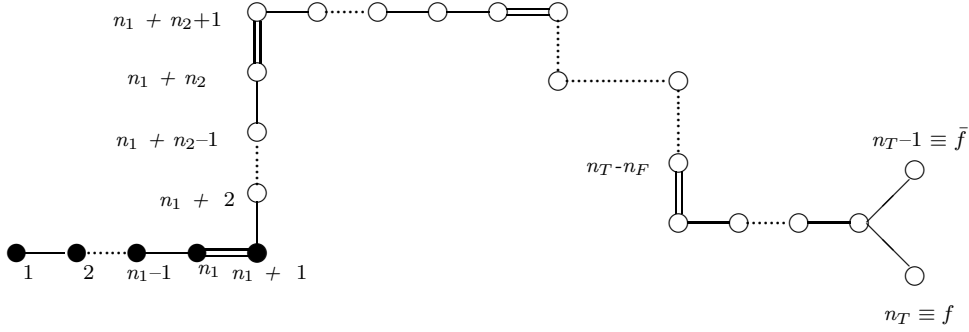
\begin{figure}[htbp]
\begin{center}
\begin{picture}(130,100)(0,0)
\put(-25,60) {\usebox{\Bn}}
\put(-20,20){\parbox{130mm}{\caption{\label{fig1} \protect {\sts
The TBA graph associated to the SG model in  a rational point:
there are $n_1$ black nodes associated to the breathers , the
node $n_1$ corresponds to the soliton, all the other nodes  are
magnons. The quantity $c_{i,j}$ associated to a pair of nodes
$\{i,  j \}$ is equal to  $0$ if the nodes are not directly connected by
a link,  to $1$ if the link is horizontal and  $-1$ if
vertical. The links on the bifurcation are  horizontal if
 $F$ is even  or vertical if $F$ is odd. The double-link defines the
concatenation of an horizontal (vertical)  with a vertical (horizontal)
sub-diagram  with a change of the shift  $S_{i}$.  }}}}
\end{picture}
\end{center}
\end{figure}

The SG theory provides  the simplest example of an exactly solvable
relativistic quantum field theory. The SG
Lagrangian is
\eq
L= { 1 \over 2 } \de_{\mu} \phi \de^{\mu} \phi + { g \sqrt{ 4 \pi} \over \beta}
\cos\lf({\beta \over \sqrt{4 \pi}} \phi \ri) \virg
\en
and its  S-matrix is the minimal $O(2)$-symmetrical solution of the
unitarity, crossing and factorization equations~\cite{zamzam}. The
spectrum of the theory consists of a soliton and a
number of soliton-antisoliton bound states with masses
\eq
M_j=2 M_{sol} \sin \lf( {j \xi \over 2} \pi \ri) \sps j=1,2,\dots <{1 \over
\xi}  \virg
\en
\eq
\xi= {\beta^2/8 \pi  \over 1-\beta^2/8 \pi} \pu
\en
In order to obtain the TBA equations we have to apply the algebraic Bethe
ansatz, and in developing this procedure  in general
in addition to the physical particles a finite number of pseudoparticles
(also known as  magnons)   describing  the colour interchange of the
soliton must be introduced.
The thermodynamic problem for the sine-Gordon model in the repulsive
region  is directly related
to those of the XXZ models solved by Takahashi and Suzuki~\cite{ts}.
For finding  the TBA
equation in both the attractive and repulsive regimes  directly from  the
SG S matrices we used  standard methods~\cite{yy,al2,ts,jf} and    the
details of this
derivation are not important in this  context.
The passing  from the TBA equation to a ``minimal'' compact expression for
the associated Y-system is quite a direct but a tedious calculation ,  our
results confirm the structure
depicted in~\cite{ta}: the sine-Gordon Y-system can be represented
as a nested structure of a concatenation of $A_{n_i}$ Dynkins diagram ending
with a $D_{n_F}$ diagram, to give a $D_{n_T}$-like Dynkin diagram.
To be more precise,  the Y-system is uniquely defined by
the simple continued-fraction~\footnote{See~\cite{cont} for
a rigorous  definition and for a list of mathematical applications
of the continued-fractions theory.} representation
for the spectral
parameter $\xi$~\footnote{ See~\cite{ts} for a classification
of the string solutions of the XXZ model using the simple
continued-fractions.}. Let us consider first the case  $n_1 \ne  0$,
\eq
\xi={p \over q-p}=\hat{\xi}(n_1,n_2, \dots , n_F ):=
 { { 1 \over n_1 + {1
\over n_2 + \dots {1 \over n_F-1}}}} \virg
\label{con}
\en
the  TBA  equation  at this  point contains $n_1$
breathers,   a soliton  and
$\sum_{i=2}^{F} {n_i}$ magnons for a total of $n_{T}=\sum_{i=1}^{F} {n_i}$
(pseudo)particles . Defining  the shifts
\eq
s_1=\imath \pi { \xi_1  \over 2 }
 ~~, ~~
s_2=\imath \pi {\xi_1 \xi_2 \over 2} ~~,~~ \dots ~~,~~ s_F=\imath \pi {
\xi_1 \xi_2 \dots \xi_F \over 2} \equiv {\imath \pi \over 2 q-2 p} ~~, \en
with
\eq
\xi_i=\hat{\xi}(n_i,n_{i+1}, \dots , n_F)
\virg
\en
we associate to  any
(pseudo)particle a node in a $D_{n_T}$ Dynkin-like  diagram.
In ~{\em figure}~\ref{fig1}
a graphical
representation, a
generalization of
those in~\cite{ta} for the SG at $\xi= {2 n+1 \over 2} $ ,
is introduced for the readers convenience;
the first
$n_1$-nodes on the tail correspond to the breathers and the node $n_1+1$
to the soliton, the others are the magnons. In the following we shall indicate
as $f \equiv n_T$ and
$\bar{f} \equiv n_T-1$ the two nodes on the bifurcation of the $D_{n_T}$
diagram.
Defining for  any node a shift
 \eq \CS_i= s_a \sps \sps \sum_{k=1}^{a-1}
n_k < i \le \sum_{j=k}^{a} n_k
\virg
\en
and for  any pair  of nodes $\{ i, j \}$ an exponent
\eq
c_{i,j}~=~c_{j,i}= (-1)^{a-1} \hspace{1 cm} \virg \hspace{1 cm}
 \sum_{k=1}^{a-1} n_k <
i,j \le 1+\sum_{k=1}^{a} n_k \virg
\label{sign}
\en
with $c_{i,j}=0$ if $i$ and $j$ are not adjacent.
The Y-system for the nodes in the set
$\{n_1,n_1+n_2, \dots , n_T-n_F-n_{F-1}\}$ , that is $k=\sum_{i=1}^a n_i$ and
$ a < F-1$,  is
\eq
\ba{c}
\ds{
Y_k \lf(\th+\CS_k\ri)~Y_k \lf(\th-\CS_k\ri)~=~(1+Y_{k-1}(\th)^{c_{k,k-1}}
)^{c_{k,k-1}} } \acc
\ds{
(1+Y_{k+n_{a+1}+1}(\th)^{\tilde{c}_{k}
})^{\tilde{c}_{k}}
} \acc
\ds{
\prod_{j=k+1}^{k+n_{a+1}}
(1+Y_{j}(\th + (k+n_{a+1}-j)
\CS_j+\CS_{k+n_{a+1}+1})^{\tilde{c}_{k}})^{\tilde{c}_{k}}
\label{y0}
} \acc
\ds{
\prod_{j=k+1}^{k+n_{a+1}}
(1+Y_{j}(\th - (k+n_{a+1}-j)
\CS_j-\CS_{k+n_{a+1}+1})^{\tilde{c}_{k}})^{\tilde{c}_{k}} \virg
}
\ea
\en
with $\tilde{c}_j=c_{j,j+1}$.

\vskip 1cm

\noindent
For the node $k=n_T-n_F$

\eq
\ba{c}
\ds{
Y_k \lf(\th+\CS_k\ri)~Y_k \lf(\th-\CS_k\ri)~=~
(1+Y_{k-1}(\th)^{c_{k,k-1}})^{c_{k,k-1}}
} \acc
\ds{
(1+Y_{f}(\th)^{\tilde{c}_k })^{\tilde{c}_k}~
(1+Y_{\bar{f}}(\th)^{\tilde{c}_k })^{\tilde{c}_k}
} \acc
\ds{
\prod_{j=k+1}^{n_T-2}  (1+Y_{j}(\th+(n_T-1-j)
\CS_j)^{\tilde{c}_k})^{\tilde{c}_{k}}
}\acc
\ds{
\prod_{j=k+1}^{n_T-2}
(1+Y_{j}(\th-(n_T-1-j) \CS_j)^{\tilde{c}_k})^{\tilde{c}_{k}} \virg
}
\ea
\en
for all the other nodes
\eq
Y_i\lf(\th+\CS_i \ri)~Y_i\lf(\th-\CS_i \ri)~=~ \prod_{j \in
adj}(1+Y_j^{c_{j,i}}(\th))^{c_{j,i}}  \pu
\label{y1}
\en
In eq.(\ref{y1}) the sum  runs over all adjacent nodes of the  $D_{n_T}$
Dynkin diagram.
Finally the Y-systems  at
$\bar{\xi}={1 / \xi}$  in the repulsive region can be obtained
from those at  $\xi$ by the change  $c_{i,j}  \fr -c_{i,j}$
and  $s_i \fr s_i/ \xi $; this is the so-called
TBA-duality in the sine-Gordon model.

\noindent
The reduced Y-system can be obtained as limit~\footnote{ The steps
involved in the deduction of this result are formally the same described
in~\cite{fen}.}
\eq
Y_{f}(\th) \fr -1 \hspace{ 1 cm} \virg \hspace{ 1 cm} Y_{\bar{f}}(\th)
\fr -1  \virg
\label{dec}
\en
of the above equations , this also forces $Y_{n_T-2} \fr 0$ or
$Y_{n_T-2} \fr \infty $.
In taking the limit~(\ref{dec}) we  substitute  the equation for the
node $n_T-2$ in the equation for the node $n_T-n_F$.
Considering that now
the last three nodes decouple,
equations~(\ref{y0},\ref{y1})   continue to hold.
The only equation that formally changes is that for  the node $k=n_T-n_F$
\eq
\ba{c}
\ds{
Y_k\lf(\th+\CS_k\ri)~Y_k \lf(\th-\CS_k\ri)~=~
(1+Y_{k-1}(\th)^{c_{k,k-1} })^{c_{k,k-1}}
} \acc
\ds{
(1+Y_{n_T-3}(\th)^{-\tilde{c}_k })^{-\tilde{c}_k}
} \acc
\ds{
\prod_{j=k+1}^{n_T-3}
(1+Y_{j}(\th+(n_T-1-j) \CS_j)^{\tilde{c}_k})^{\tilde{c}_k}
} \acc
\ds{
\prod_{j=k+1}^{n_T-3}
(1+Y_{j}(\th-(n_T-1-j) \CS_j)^{\tilde{c}_k})^{\tilde{c}_k}
}
\ea
\en

using the procedure of~\cite{rtv,fein,ta} the resulting Y-system can be
used to  describe all the
minimal reduced theories of the $Z_n$ supersymmetric-sine-Gordon models, for
instance
the case $n=2$ corresponds to the minimal $N=1$ supersymmetric series.
As usual
the  full TBA equations  can be obtained via standard Fourier
transformations.

\section{New functional dilogarithm equations}

The Rogers dilogarithm~\cite{lewin} $L(x)$ with $0\le x\le 1$ is the
unique function defined by
the integral representation
\eq
L(x)= - { 1 \over 2} \int_0^x dy \lf[ {\log(y) \over 1-y} +
{\log(1-y) \over y} \ri] \virg
\en
it  is three times differentiable and satisfies the  Euler functional equation
\eq
L(x)+L(1-x)=L(1) \virg
\label{eu}
\en
as well the
five term relationship  known also as the Spence and Abel functional
equation \eq
L(x)+L(1-xy)+L(y) +L \lf( \frac{1-y}{1-xy}\ri) +
L\lf(\frac{1-x}{1-xy}\ri)=3\,L(1) \pu
\label{abel}
\en
Together with the periodicity properties that can be numerically
verified with high precision
\eq
\ba{c}
\ds{ Y_i \lf( \th+ \imath \pi {1 + \xi \over 2}  \ri)=Y_i (\th)} \virg \acc
\ds{ Y_{f} \lf( \th+ \imath \pi { 1 + \xi \over 2} \ri)=Y_{\bar{f}} (\th)}
\virg
\acc \ds{ Y_{\bar{f}} \lf( \th+ \imath \pi { 1 + \xi \over 2} \ri)=Y_{f}
(\th)} \virg \ea
\en
we have all the  basic ingredients in
order to construct our  FDE.
Following~\cite{gt}
we  introduce   the function
\eq
H(x)= L\lf({x\over 1+x} \ri) / L(1) \pu
\label{hh}
\en
and defining   the quantities
\eq
Y_i \lf( \th + \imath \pi { m  \over {2 q - 2 p} }\ri)=\Y_i(m) \virg
\en
with  an appropriate number of  initial conditions we find  the  following
identities \footnote{ $ \xi < 1$ is implicit and the  equations
for $ \bar{\xi} > {1 \over \xi}  $ can be  obtained by using the Euler
formula. }
 \eq
\D_{p,q}(SG) ::= \sps  \sum_{i=1}^{n_T} \sum_{m=0}^{2 q-1}  H(\Y_i(m))=
 2 q \lf( \sum_{i=1}^{[(F+1)/2]}  n_{2 i-1} + (-1)^F \ri)
\label{dilog}
\en
for the SG TBA. F is the integer defined in eq. (\ref{con}) and $[(F+1)/2]$
is the integer part of $(F+1)/2$.
For the RSG
\eq
\D_{p,q}(RSG)::= \sum_{i=1}^{n_T} \sum_{m=0}^{2 q-1}  H(\Y_i(m))= \N
\label{dilog1}
\en
and
\eq
\N = 2 q \lf( \sum_{i=1}^{F} 6 { (-1)^{i+1} \over p_i q_i}
+ \sum_{i=1}^{[(F+1)/2]}n_{2i-1} - \lf({3 \over 2} - (-1)^F { 7 \over
2}\ri)\ri)
\label{NNN}
\en
with $p_1 = p , q_1=q$ but
\eq
{ p_i \over q_i -p_i} = \hat{ \xi}(n_i-1 , n_{i+1} , \dots, n_F  )
\virg
\en
for $i>1$.
The quantity $\N$ defined in eq.~(\ref{NNN}) is  a positive integer
and in the first simplest cases it takes the following values
\eq
 \sps \N= 2 (n_F-3)(n_F-2)
\en
for $F=1$,
\eq
 \sps \N= 2(2 n_F-8n_1+3n_Fn_1-{n_1}^2+n_F {n_1}^2-6)
\en
for $F=2$, and
\eq
\ba{c}
\ds{ \N= 2(6 - 5 n_F + {n_F}^2 - 5 n_1 + 2 n_F n_1 + {n_1}^2 + 11 n_2} \acc
\ds{ - 6 n_F n_2 + {n_F}^2 n_2 + 10 n_1 n_2 - 5 n_F n_1 n_2} \acc
\ds{ + {n_F}^2 n_1 n_2 - {n_1}^2 n_2 + n_F  n_2 {n_1}^2 )}\acc
\ea
\en
for $F=3$.
Equations~(\ref{dilog}) and~(\ref{dilog1}) are  the main result of this
letter. In general in the sums~(\ref{dilog},\ref{dilog1} ) there are two
identical independent FDE.  It is possible to
remove one solution with an identical  choice of the initial
conditions and dividing by two the final  result. In the following we
will assume that this procedure is applied.
We have the following identification with the  theories in~\cite{gt}:
$\D_{1,p}(SG)  \equiv \D( D_p \diamond A_1 )$ ,
$\D_{1,p}(RSG)  \equiv \D( A_{p-3} \diamond A_1 )$ and
$\D_{2,2 p+3}(RSG) \equiv \D( T_p \diamond A_1 )$,  the first low rank
new equation is the $\D_{3,7}(RSG)$ with four free parameters. The
functional equations are
\[
\Y(m+3) \Y(m-3)= (1+\CZ(m+2)) (1+\CZ(m-2)) \lf(1 + {1 \over \CZ(m)}
\ri)^{-1}
\]
\eq
\CZ(m+1) \CZ(m-1)= (1+\Y(m)) \virg
\en
the with   initial conditions  $\{ \Y(0) = x~,~ \Y(2) = y~,~ \Y(4) =
z~,~ \CZ(1) = k \}$ we find

\[ \D_{3,7}(RSG) \sps  ::= \sps
   H(k)  + H(x) + H(y) + H(z) +
\]
\[
  H \lf( {1 + y \over k}  \ri) + H \lf( {k (1 + z) \over 1 + y}
      \ri) + H \lf( {1 + x \over k } \ri) +
\]
\[
 H \lf( { (1 + y) (1 + 2 k + k^2 + x + k x + y + k y + x y + k z +
 k^2 z) \over    k x (1 + k + y) (1 + z)} \ri)+
\]
\eq
H \lf( {(1 + k) (1 + x) (1 + k + y) (1 + z) \over
    y (1 + 2 k + k^2 + x + k x + y + k y + x y + k z + k^2 z)} \ri)+
\en
\[
 H \lf( { 1 + 2 k + k^2 + x + k x + y + k y + x y + k z + k^2 z  \over
    (1 + k) (1 + x) z } \ri)+
\]
\[
H \lf( { (1 + k) (1 + k + y + k z)  \over  x (1 + k + y)} \ri)
+ H \lf( {1 + k + x + k x + y + k y + x y \over k x y} \ri)+
\]
\[
H \lf( {1 + k + y + z + k z \over  y z} \ri)
   + H \lf( {(1 + k + x) (1 + k + y)  \over k z (1 + k) } \ri) =  9 \pu
\]
\section{ Comments  and conclusions}
Finally let us make  some comments on  the   TBA-duality
characterized in~\cite{ta} and  connecting two
theories in different regimes in  the SG and in the RSG models.
First we note that    this  duality-like relation  is  the
off-critical  version  of  the symmetry
\eq
 { p \over  q} \leftrightarrow  {q-p \over  q}
\virg
\en
of the counting solution  for the XXZ chain~\cite{bc},
and that
in the   RSOS models this  transformation  connect the critical line in
the regime III/IV to that in the regime I/II.
An important generalization of the L-state  RSOS are the RSOS(P,P)
models
obtained by the fusion of $P \times P$ blocks of face weights together.
The continuum limits of these theories in the regime III/IV are
( see ref. \cite{bz} and references therein) the
$ SU(2)_P \times SU(2)_{L-1-P} / SU(2)_{L-1}$ GKO constructions, but
 in the regime I/II all these theories have as same continuum
limit   the $Z_{L-1}$ models.  This multi-to-one correspondence  has an
off-critical
analogue: the $A_1 \diamond A_{L-2}$  Y-systems  describe all the
perturbations $SU(2)_P \times SU(2)_{L-1-P}/ SU(2)_{L-1}
+\phi^{ id,id }_{adj}$, but the $A_{L-2} \diamond A_1$
describe only  the $Z_{L-1} +\ep_1 $ model.
It is not completely  clear to us
if this  multi-to-one correspondence is also verified in the  general
case  $L-P={ p \over q -p} -1$ .
Obviously in  this   case  instead of the $Z_{L-1}$ the common theory
should be the theory $M_{q-p,q}$.
Another  interesting point consists in the existence of integrable
perturbations connecting the theories in these two regimes;
in~\cite{faza}  non perturbative
fluxes between the $Z_N$ models and the unitary series have been  proposed
\eq
Z_{L-1} + \psi_1 \bar{\psi_1}+ \psi_1^{\dagger} \bar{\psi}_1^{\dagger}
\fr M_{L,L+1}
+\phi_{3,1}+ \mbox{higher order terms} \pu
\label{zzn}
\en
More  recently in~\cite{rst} also  integrable perturbations  in the
non-unitary models generated by
the operator  $\phi_{2,1}$  have  been conjectured; their generalizations
imply  a RG structure of the kind
\eq
M_{p,q}+\phi_{2,1} \fr  M_{q-p,q}+ \phi_{1,5} +\mbox{higher order terms}
\virg
 \label{pp}
\en
so  exactly connecting the ($ P = 1$)   RSOS models related by the
TBA-duality!
Naively speaking we also note that in at least  two cases we can find
fluxes connecting  theories in one regime with   theories in
the other. The first example is  the $Z_4$ model,
there exist   a massless unitary trajectory
into the $M_{5,6}$ model of the kind~(\ref{zzn}),  but  also a second
possibility exists,
that is, perturbing the theory with a imaginary coupling constant~\cite{fsz}:
in this case due to the   ``equivalence''~\cite{zn}  with  the sine-Gordon
model at $\xi= 2 $ we find   as IR limit the
theory $SU(2)_2 \times SU(2)_2 / SU(2)_4$.
The  simplest non unitary theory with at least two particles is
the models $M_{2,7}$, in this case the  TBA-duality  fixes the
correspondence
\eq
M_{2,7} \leftrightarrow \{ M_{5,7}~~,~~M_{7,12} \}  \virg
\en
the massless perturbations  connecting the theory $M_{5,7}$ versus the
theory $M_{2,7}$ is a  consequence of the conjecture~(\ref{pp}).
The flow  connecting the theory  $M_{7,12}+\phi_{1,4}$ with the theory
$M_{2,7} + T \bar{T} + \mbox{h.o.t.} $
can be found in the family of TBA of ~\cite{rt} and associated to the
$ T_2\diamond A_2$ theory.
It is an interesting open problem  to find other examples  of this kind of
non-perturbative RG fluxes in the RSOS(P,P) models.
Finally we note that a generalization
of the  two-graphs tensorial product  introduced
in~\cite{ra}, should permit to find the TBA systems for many models in the
families
 $G_k \times G_l \over G_{k+l}$ with
  $k$ and $l$ particular rational numbers.

\vskip .6cm

\noindent
{\bf Acknowledgements} --
We are greatly indebted to P.Dorey , F.Gliozzi and F.Ravanini for a lot of
very useful discussion and help.
We also are grateful to R. de Jeu and
 K.E.Thompson  for useful discussions and  the
Mathematics Department of Durham University for the kind hospitality.

\end{document}